\documentclass[
]{ceurart}

\usepackage{listings}

\usepackage{algorithm}
\usepackage{algpseudocode}

\begin{document}

\copyrightyear{2024}
\copyrightclause{Copyright for this paper by its authors.
  Use permitted under Creative Commons License Attribution 4.0
  International (CC BY 4.0).}

\conference{Information Retrieval's Role in RAG Systems (IR-RAG) - 2024}

\title{RAGSys: Item-Cold-Start Recommender as RAG System}

\author[1]{Emile Contal}[%
email=emile@crossingminds.com,
]
\cormark[1]
\fnmark[1]
\address[1]{Crossing Minds Inc, San Francisco, USA}

\author[1]{Garrin McGoldrick}[%
email=garrin.mcgoldrick@crossingminds.com,
]
\fnmark[1]

\cortext[1]{Corresponding author.}
\fntext[1]{Both authors contributed equally to this research.}

\begin{abstract}
Large Language Models (LLM) hold immense promise for real-world applications, but their generic knowledge often falls short of domain-specific needs. Fine-tuning, a common approach, can suffer from catastrophic forgetting and hinder generalizability. In-Context Learning (ICL) offers an alternative, which can leverage Retrieval-Augmented Generation (RAG) to provide LLMs with relevant demonstrations for few-shot learning tasks. This paper explores the desired qualities of a demonstration retrieval system for ICL. We argue that ICL retrieval in this context resembles item-cold-start recommender systems, prioritizing discovery and maximizing information gain over strict relevance. We propose a novel evaluation method that measures the LLM's subsequent performance on NLP tasks, eliminating the need for subjective diversity scores. Our findings demonstrate the critical role of diversity and quality bias in retrieved demonstrations for effective ICL, and highlight the potential of recommender system techniques in this domain.
\end{abstract}

\begin{keywords}
  Recommender systems \sep
  Information Retrieval \sep
  Large Language Models \sep
  Few-Shot Learning \sep
  In-Context-Learning
\end{keywords}

\maketitle

\section{Introduction}

Large Language Models (LLMs) have emerged as a powerful tool for natural language processing, demonstrating remarkable abilities in areas like text completion, summarization, and question answering \cite{brownNEURIPS2020}. One of their most intriguing capabilities is their potential to learn "common sense" –  general knowledge about the world that allows them to reason and make inferences beyond the literal meaning of text. This has fueled excitement about the possibility of achieving zero-shot learning, where LLMs can solve unseen problems without any prior training on specific tasks \cite{kojimaNEURIPS2022}.

However, a crucial distinction exists between generic public knowledge and the specific private knowledge required for most real-world use cases. While LLMs excel at generic text completion or chat-like interactions, practical applications often demand solving specific and repeatable downstream tasks within a particular domain \cite{HuICLR2022}. This typically necessitates knowledge specific to a business or organization, such as understanding internal processes, up-to-date product details, or customer behavior.

Fine-tuning, a technique where LLMs are trained on large datasets tailored to the target task, offers a path towards adapting LLMs to these domain-specific needs. Yet, fine-tuning presents significant challenges. When trained on tasks-specific data, LLMs tend to forget knowledge and skills gained in the initial training, a phenomenon referred to as Catastrophic Forgetting \cite{Luo2023AnES}. Consequently, a fine-tuned LLM loses some of its ability to generalize to novel examples that aren't well represented in its fine-tuning training data. Moreover, while fine-tuning allows an LLM to memorize task-specific information, it doesn't necessarily allow the LLM to reason about that information \cite{Berglund2023TheRC}. As a final consideration, keeping LLMs constantly up-to-date using fine-tuning can be infeasible, especially for domains with frequently changing information like e-commerce product inventory, whereas it is easy to update a database in real-time from which information is retrieved.

As an alternative to fine-tuning, In-Context Learning (ICL) offers a promising approach for leveraging LLMs in scenarios with limited data. This approach exploits the demonstrated ability of LLMs for "meta-learning" – essentially, learning how to learn.
In \cite{BaiNEURIPS2023}, the authors prove the capacity of LLMs to effectively ingest in-context training data points and solve statistical optimization problems such as gradient descent.
ICL enables practitioners to leverage Retrieval Augmented Generation (RAG),
that is enriching the input prompt by information that is retrieved in real-time \cite{RamTACL2023}.
We refer to \cite{QingxiuARXIV2023} for a recent survey on ICL.

This paper focuses on few-shot learning and the retrieval of relevant demonstrations for this process, where a demonstration is some text which is included in the LLM's context to demonstrate how the LLM should formulate correct answers. Few-shot learning presents a well-structured problem, allowing us to evaluate the quality of the retrieval algorithm using established classification metrics. Crucially, we show that enriching a language model with a few-shot example retriever offers a powerful method to achieve fine-tuning-like behavior, steering the output of the LLM towards the desired outcome even with limited data.
Interestingly, increasing the context size in prompts beyond a certain point yields diminishing returns. The most impactful information resides within a relatively small set of well-chosen demonstrations, rather than overloading the prompt with vast amounts of data \cite{MachlabARXIV2024}. This highlights the importance of effective retrieval strategies, transforming $k$-shot learning into a top-$k$ information retrieval problem at its core.

Building upon this concept, this paper identifies desirable properties for a RAG system under the framework of few-shot learning. We demonstrate that state-of-the-art retrieval systems in this context resemble item-cold-start recommender systems. Unlike exact search algorithms that prioritize precision and recall, our focus is on discovery, by maximizing the set of collective information gain from the retrieved demonstrations. This necessitates solving various trade-offs between query relevance, quality scoring, as well as diversity algorithms to ensure a variety of informative examples are surfaced.
Furthermore, we propose a method for evaluating RAG system performance through the subsequent performance of the enriched LLM on established NLP tasks like question answering or text generation. This methodology offers a valuable approach to directly assessing diversity and quality-based retrieval systems, which removes the need to define a subjective diversity score, a historically challenging aspect of evaluating such systems in academic settings \cite{clarkeSIGIR2008}.

To summarize, in this paper we study the impact of diversity and quality bias in retrieving demonstrations for ICL.
We start by reviewing the use of diversity and other biases in both ICL and Information Retrieval works.
We then propose a method for evaluating the performance of different retrieval algorithms. 
Then we present experiments and results demonstrating the impact of diversity and quality bias on an LLM's ability to generate correct answers. 
Finally we discuss the applicability of state-of-the-art ICL retrieval algorithms in real-world setting,
and show that recommendation engines offer a better solution than semantic search engines.

\section{Related Work}

This paper sits at the intersection of two distinct, but increasingly intertwined, research areas: In-Context Learning for Large Language Models and Information Retrieval. While ICL focuses on enabling LLMs to learn from carefully selected contextual information, IR deals with retrieving relevant information from document collections. Our work leverages concepts from both fields to address the challenge of few-shot learning with LLMs.

\subsection{Few-Shot In-Context Learning with Retrieval Augmented Generation}

Within the context of RAG, few-shot learning can be defined as a specific scenario where the "documents" retrieved are actually "examples" used to guide the LLM. These examples can also be referred to interchangeably as "In-Context Examples" (ICE) or "demonstrations". The importance of ICL in achieving state-of-the-art LLM performance is undeniable, with its ubiquitous presence in top benchmarks across various domains. Consequently, ICL research is a rapidly evolving field with numerous proposed algorithms.

\subsubsection{Pure Diversity}
Several noteworthy ICL approaches have emerged that address the challenge of retrieving informative examples for few-shot learning. Some methods only promote the diversity of the demonstrations, like in \cite{yuICLR2023} where the authors utilize $k$-means clustering in a dense embeddings space to achieve diversity. By applying $k$-means to the sentence embeddings of the demonstrations, this approach ensures that the retrieved examples cover a variety of semantic spaces, inherently increasing the mutual information of the retrieved set, but without taking into account the relevancy to the query.

\subsubsection{Pure Quality}
Other approaches focuses on identifying examples where the LLM exhibits low token-level uncertainty. In \cite{gonenEMNLP2023} the authors analyze token probabilities within candidate 0-shot prompts. By prioritizing prompts where the LLM has the highest generation likelihood (low perplexity), this approach aims to select examples that hold the potential for significant learning gains for the LLM.
The intuition that the authors give is that a prompt that is more expected by the LLM is more likely to help it extracting the relevant information.
Accessing the per-token probabilities for all examples incurs a significant compute cost, but can be pre-computed as they do not depend on the query.

\subsubsection{Pure Relevance}
Notably, a connection can be drawn between traditional full-text search algorithms and pure relevance approaches. In \cite{agrawalFINDINGSACL2023} the authors use BM25 \cite{robertsonNIST1995}, a well-established retrieval ranking function commonly used in information retrieval tasks. This approach essentially leverages the strengths of BM25 in identifying examples with terms highly similar to the query, to select the most relevant examples for the specific task at hand. This strategy ensures the retrieved examples are topically relevant to the task while potentially introducing some variation in the specific phrasing or wording used.

Finally, neural ranking, one of the most commonly used ICL approach, typically yielding superior results \cite{rubinNACL2022, wangARXIV2022, liACL2023, yeICML2023, wangEACL2024}, is maximizing similarity in a dense embeddings space. These methods, like \verb|KATE| \cite{liuDEELIO2022}, utilize $k$-Nearest Neighbors (\verb|kNN|) search using the cosine distance of sentence embeddings to retrieve the examples most semantically similar to the prompt. Scaling this method leverages vector search algorithms, commonly used in large-scale information retrieval tasks, where efficient retrieval of semantically similar documents is crucial.

While general purpose pre-trained embeddings like \verb|BERT| \cite{devlinNACL2019} form a strong baseline, learning specific embeddings for retrieval, and in particular ICL, is a very active area of research. In \cite{khattabSIGIR2020} the authors build upon \verb|BERT| and introduce \verb|ColBERT| that improves the retrieval performances by an order of magnitude.
Other embeddings model have been proposed for ICL retrieval in \cite{wangARXIV2022} and \cite{liACL2023}.
Some authors also explored training full language models \cite{rubinNACL2022}, as well as LLM \cite{wangEACL2024}, showing further improvements compared to traditional embedding-based approaches.
While leading to superior results, these supervised neural ranking models for learning-to-rank necessitate orders of magnitude more training examples data, that is typically not available to practitioners. In addition, without any explicit metric space such as dense embeddings, efficient retrieval indexing such as \cite{beygelzimerICML2006} cannot be used.

\subsubsection{Diversity/Relevance Trade-off}
While both relevance and diversity are crucial for effective few-shot learning, methods yielding the best ICL results combine these two paradigms rather than prioritizing one over the other. This is achieved by maximizing a carefully balanced trade-off between semantic similarity to the prompt and diversity of the retrieved examples. Unsupervised techniques can be adapted to prioritize the selection of examples that are both relevant to the prompt and dissimilar to each other.
In \cite{levyACL2023} the authors introduce a greedy method to select relevant demonstrations while ensuring enough coverage. They define specific coverage strategies adapted to the problem of program generation. 
In \cite{suICLR2023} the authors employs an active learning setting, where a voting algorithm selects a set of examples penalizing the top-$k$ closest from already selected ones, using cosine distance in an embedding space.

The most popular unsupervised approach for achieving this balance between relevance and diversity is Maximal Marginal Relevance (MMR). MMR retrieves a set of examples by iteratively selecting the example that maximizes a linear combination of the scores of relevance to the prompt and dissimilarity to the previously retrieved examples.
It was analyzed in \cite{yeFINDINGSACL2023} for ICL and was shown to outperform simpler methods.
Alternatively to MMR, Determinantal Point Processes (DPP) has been used in \cite{yeICML2023} to optimize the joint information of the selected $k$ examples. However, exactly solving the DPP optimization being \verb|NP|-hard, hence the authors also employs greedy maximization.

\subsection{Diversity in Information Retrieval and Online Learning}

The concept of diversity in information retrieval has been a long-running topic of research.
In this section we propose a short review of the use of diversity in the IR literature and related domains.

\subsubsection{In Information Retrieval}
The MMR algorithm was analyzed in \cite{carbonellSIGIR1998}, and compared against other approached like \verb|KL|-divergence optimization in \cite{zhaiSIGIR2003}.
Pure retrieval algorithms typically optimize for recall, not information maximization.
Agreeing on a diversity objective function remains a challenge.
Diversity is sometimes introduced as a heuristic to cover possible interpretations of the same query,
instead of minimizing information overlap from near-duplicated results.
In \cite{clarkeSIGIR2008} the authors leverage a concept of {\em information nuggets} with documents to estimate the redundancy of the set of retrieved documents.
Topic modeling is also employed, such as \cite{agrawalWSDM2009} that uses a taxonomy of categories labelling the documents and the queries.
The desired properties of diverse retrieval are furthermore characterized in \cite{gollapudiWWW2009}.
A various set of similarity methods and diversification algorithms are analyzed in \cite{vieiraICDE2011} on sparse features vectors.
Among diversity evaluation methods based on topic modelling, three notable criteria used in the \verb|TREC| Diversity track \cite{clarkeTREC2009}, \verb|ERR-IA| \cite{chapelleCIKM2009}, $\alpha$-\verb|nDCG|@$k$ \cite{clarkeSIGIR2008}, and \verb|NRBP| \cite{clarkeICTIR2009}, are compared in \cite{clarkeWSDM2011}.

\subsubsection{In Recommender Systems}
Within IR, the recommender system literature brings an additional point-of-view on studying diversity in retrieval, by focusing on the benefit of diverse results for a user, instead of evaluating methods against a potentially arbitrary relevance/diversity trade-off.
The difficulty of evaluating the impact of diversity, and the necessity for large scale real-world recommendation studies has been explored in \cite{zieglerWWW2005}.
In \cite{vargasRECSYS2011} and \cite{varagsSIGIR2014} the authors model the user behavior conditioned on the set of retrieved items.
In \cite{zhengWWW2021} the authors improve the diversity versus relevance trade-off in recommender systems by directly learning a ranking model that favor diversity, instead of only applying diversity re-ranking methods.

\subsubsection{In Online Learning}
Learning a trade-off between relevancy and diversity also naturally occurs in related online frameworks such as active learning, multi-armed bandits and Bayesian optimization.
In \cite{radlinskiICML2008} the authors modify a learning-to-rank algorithm from users feedback, to inherently learn diverse rankings and demonstrate a positive impact on the original relevance metric.
Other approaches such as \cite{zhuSIGIR2014} also introduce diversity in learning-to-rank algorithms while preserving the offline settings, but then are limited to evaluate using direct diversity measures.

Within batch-mode Bayesian optimization, in \cite{desautelsJMLR2014} and \cite{contalECMLPKDD2013} the authors analyze two greedy exploration/exploitation algorithms to select the next batch of items maximizing the cumulative reward. Like with recommender systems, these online settings exemplify the theoretical and empirical importance of diversifying the selected set of items despite the true objective function only including pure relevance, the cumulative reward.

\subsection{Quality Biases in Information Retrieval}

Complementing the discussion on diversity in information retrieval, quality bias also plays a crucial role in effective retrieval. Quality bias refers to the prioritization of documents or examples considered to be more reliable or informative within the retrieved set. Incorporating quality considerations into retrieval algorithms can significantly improve standard unbiased IR metrics.

Several approaches have been explored to address quality bias in pure IR tasks. These can be broadly categorized into content-based and graph-based methods.

\subsubsection{Content-Based Quality Biases}
Content-based methods leverage existing signals inside the documents themselves to identify potentially lower-quality content. Examples include spam detection scores developed in works like \cite{ntoulasWWW2006} and \cite{cormackIR2011}. By incorporating such scores during retrieval, the system can prioritize higher quality documents.
More sophisticated content-based approaches don't limit at spam classification, but extract more generic quality features the content of documents. Works like \cite{benderskyWSDM2011} explore features such as stop-word statistics or entropy of the documents to generate quality scores. The authors demonstrate that biasing standard retrieval using these features leads to improved retrieval effectiveness even using unbiased IR metrics like \verb|nDCG|.

\subsubsection{Graph-Based Quality Biases}
Instead of relying on the content itself, graph-based algorithms inherently capture implicit quality signals within their ranking model. \verb|PageRank|, a seminal algorithm for web search ranking introduced in \cite{brinCISDN1998}, exemplifies this approach. \verb|PageRank| leverages the links structure between web articles to assign higher importance to web pages that are linked to by other high-quality pages. This process implicitly prioritizes documents with a higher perceived quality based on the quality of their in-links.

\subsubsection{Connections to Recommender Systems}
Interestingly, the concept of inherent quality bias in graph-based IR approaches resembles collaborative filtering techniques employed in recommender systems. In an analogous manner to learning-to-rank on a (item, item) graph, collaborative filtering addresses learning-to-rank on a bipartite (user, item) graph. In this way, collaborative filtering also implicitly learns a trade-off between item similarity and popularity, favoring items that are both similar to the user's past preferences and also generally well-received by other users.

\section{Methodology}

We propose to frame the ICL problem as an item-cold-start recommendation problem, where the query is an unseen item, and the objective is to retrieve from the pool of candidate few-shot demonstrations a set of items maximizing the cumulative reward to the user (the LLM). In this case, the reward is a measure of how much the retrieved items increase the probability that the LLM generates a correct answer. A solution to this optimization problem requires not only relevance, but also diversity and quality in the retrieved items, such that the amount of useful information presented to the LLM in the context is maximized.

Further, we propose to measure the impact of diversity on the retrieved items by directly calculating the probability of the LLM generating a correct answer given the context items. This is in contrast to a typical retrieval context where the retriever is evaluated by calculating some metric relating to the accuracy and recall of documents most similar to the query. In such a setting, it is typical to add a term to the metric which measures the diversity of the retrieved documents to promote more diverse retrievers, knowing that diversity improves the reward to the user but without having an explicit model connecting diversity to the user's reward. In the case of retrieving demonstrations for inclusion in an LLM's context, we can directly measure the impact of diversity on the LLM's reward by calculating the probability of the LLM generating a correct answer.

\subsection{Problem Statement}

Consider an answer $a$ that should be generated by an LLM in response to a query $q$. The query can be a simple question such as "Who was the first man to walk on the moon?", or more general message such as "I'd like to find red shoes". The answer could take on many forms, such a factual response "Neil Armstrong", a clarifying question "What size and style of shoe are you looking for?", a JSON payload to send to an API "\texttt{\{"search\_terms":["red","shoe"]\}}" etc. 

Consider a set of demonstrations $\mathcal{D}$, where each demonstration is a pair $(q, a)$ containing a query $q$ and correct answer $a$, or a triple $(q, a, \bar{a})$ which additionally contains an incorrect answer $\bar{a}$. Datasets under this later triplet form are commonly used in Contrastive Learning approaches. We call $C$, a subset of demonstrations retrieved from $\mathcal{D}$, the context.

$$
C \subset \mathcal{D} = \big\{ (q_i, a_i, \bar{a}_i), \dots, (q_n, a_n, \bar{a}_n) \big\}
$$

Given an auto-regressive LLM $M$, the query $q$, and a retrieved context $C$, we define $p_M(a \mid q, C)$ the probability that $M$ generates the answer $a$.
In practice, the tokens of the examples from the context $C$ are appended to the tokens of the query $q$, using prompt formatting techniques that may be optimized to a specific LLM.


Putting it all together, for an unseen query $q$ and unseen correct answer $a$, a few-shot retriever $R_{\mathcal{D}}$ must efficiently retrieve a subset of $k$ demonstrations $R_{\mathcal{D}}(q) \in \mathcal{D}^k$ 
such that $p_M(a \mid q, R_{\mathcal{D}}(q))$ is maximized.

\subsection{Evaluation}

Consider the probability of generating the correct answer $a$ given an empty context $p_M(a \mid q)$. We are interested in evaluating how much the context $C$ increases the probability of the correct answer $a$. That is, we want a metric which is related to difference of $p_M(a \mid q, C)$ and $p_M(a \mid q)$.

In a pure retrieval setting, we would be interested in finding the context $C$ which contains the $k$ demonstrations that are most similar to the $q$. And we could argue that if there exists a smooth function $f: q \to a$ which maps a query to its correct answer, then by retrieving the demonstrations whose queries are nearest to $q$, we should also be retrieving the answers which are closest to $a$, and this should help the language model $M$ generate the correct answer $a$.

However, it is doubtful that the space in which $q$ is compared to the demonstrations is one in which the function $f: q \to a$ is smooth, so it is not necessarily true that the retrieved answers are closest to $a$. Nor is it necessarily true that $p_M(a \mid q, C)$ is maximized when $C$ contains those answers closest to $a$. Consider that the answer $a$ might depend on some information which isn't contained in $a$ or any nearby answer. 

Therefore, we prefer to measure $p_M(a \mid q, C)$ directly. In practice, given that $M$ is an auto-regressive language model, this is done by taking the product of the probability of each token generated by $M$. The model generates text sequentially by predicting one token at a time based on the previously generated tokens. Let $a = (a_1, a_2, \ldots, a_n)$ represent a sequence of tokens produced by the model. The probability of the model generating the sequence $a$ can be expressed as the joint probability of generating each token in the sequence, conditioned on the tokens that precede it. This can be mathematically represented as:
$$
p(a) = p(a_1) \cdot p(a_2 | a_1) \cdot p(a_3 | a_1, a_2) \cdots p(a_n | a_1, a_2, \ldots, a_{n-1})
$$

Thus, $p_M(a \mid q, C)$ is the product of the conditional probabilities of each token, and these probabilities are output by the LLM at inference time and are readily available in APIs serving LLMs such as the OpenAI API.

\subsubsection{Classification Metrics} \label{sec:classsification-metrics}

In binary classification, accuracy is typically used as an evaluation metric, and can be defined as:
$$
\frac{1}{|\mathcal{D}|} \sum_{(x, y, \bar{y}) \in \mathcal{D}} \mathbf{1}\big(p(y \mid x) > p(\bar{y} \mid x)\big)
$$

Where: $|\mathcal{D}|$ is the number of examples in the dataset $\mathcal{D}$; $\mathbf{1}(\cdot)$ is the indicator function that returns 1 if the condition is true and 0 otherwise; and $y$ ($\bar{y}$) is the correct (incorrect) label for example $x$.

Given a retriever $R_{\mathcal{D}}$ and a demonstration $(q, a, \bar{a}) \in \mathcal{D}$, we introduce the simplified leave-one-out notation $R(q) = R_{\mathcal{D} \setminus \{(q, a, \bar{a})\}}(q)$.
We define the metric MC1 which is related to accuracy:
$$
\text{MC1} = \frac{1}{|\mathcal{D}|} \sum_{(q, a, \bar{a}) \in \mathcal{D}} \mathbf{1}\Big( p_M\big(a \mid q, R(q)\big) > p_M\big(\bar{a} \mid q, R(q)\big) \Big)
$$

In the case that many incorrect answers are provided for each query $\mathbf{\bar{a}}$, we can extend this in the same manner as multi-class classification by requiring that the correct answer have greater probability than all the incorrect answers:
$$
\frac{1}{|\mathcal{D}|} \sum_{(q, a, \mathbf{\bar{a}}) \in \mathcal{D}} \prod_{\bar{a} \in \mathbf{\bar{a}}} \mathbf{1}\Big( p_M\big(a \mid q, R(q)\big) > p_M\big(\bar{a} \mid q, R(q)\big) \Big)
$$

We also define a metric MC2 which extends this further to the case that multiple correct answers $\mathbf{a}$ and multiple incorrect answers $\mathbf{\bar{a}}$ are provided for each query. This metric is the average number of correct answers which have greater probability than all incorrect answers.
$$
\text{MC2} = 
\frac{1}{|\mathcal{D}|} \sum_{(q, \mathbf{a}, \mathbf{\bar{a}}) \in \mathcal{D}} \frac{1}{|\mathbf{a}|} \sum_{a \in \mathbf{a}} \prod_{\bar{a} \in \mathbf{\bar{a}}} \mathbf{1}\Big( p_M\big(a \mid q, R(q)\big) > p_M\big(\bar{a} \mid q, R(q)\big) \Big)
$$

Finally, we define the related metric MC3. This metric is the ratio of probability of correct answers to the probability of incorrect answers.
$$
\text{MC3} = \frac{1}{|\mathcal{D}|} \sum_{(q, \mathbf{a}, \mathbf{\bar{a}}) \in \mathcal{D}} \frac{\sum_{a \in \mathbf{a}} p_M\big(a \mid q, R(q)\big)}{\sum_{\bar{a} \in \mathbf{\bar{a}}} p_M\big(\bar{a} \mid q, R(q)\big)}
$$

These metrics and their names follow those defined in \cite{Lin2021TruthfulQAMH}. While they are easy to interpret, these metrics are not well normalized: they don't take into account all possible correct and incorrect answers. As a result, if the sample of correct and incorrect answers have varying lengths and use of rare vocabulary tokens, these will impact the metrics.

\subsubsection{Direct Preference Optimization Metric}

We postulate that an ideal metric should obey the following properties: it should be positive when the retrieved context increases the probability of a correct answer; it should be equal in magnitude when the probability of a correct answer halves or doubles; it should relate to the probability of getting all correct answers such that if any one correct answer is impossible, the metric is minimized. Moreover, in the case that incorrect answers are provided, it should be positive when the context $C$ increases the probability of correct answers more than that of incorrect answers.

We define the DPO metric as the negative of the Direct Preference Optimization loss \cite{rafailovNEURIPS2023}, which satisfies these properties:

$$
\text{DPO} = \log \sigma \Big( \log \frac{p_M\big(a \mid q, R(q)\big)}{p_M(a \mid q)} - \log \frac{p_M\big(\bar{a} \mid q, R(q)\big)}{p_M(\bar{a} \mid q)} \Big)
$$

In the case that incorrect answers are not available, the term containing $\bar{a}$ can be omitted while preserving the aforementioned properties.

Because the metric is proportional to probability ratio $p_M(a \mid q, C) / p_M(a \mid q)$ rather than the absolute probability $p_M(a \mid q, C)$, it is invariant to the number of tokens and frequency of rare vocabulary tokens in the answer. If this were not the case, then the score for an example would get a positive (negative) bias if the correct (incorrect) answer is shorter. Similarly, the score across a set of examples would weigh examples with shorter answers more strongly.

Another aspect of the DPO metric that is worth considering is that by using this metric to optimize the retriever, we are effectively fine-tuning a model. Consider the LLM model $p_M(a \mid q)$ which assigns a probability of a generation $a$ given a prompt $q$. Now consider another model $p_M'(a \mid q')$, where $q' = R(q)$. From this perspective, we can consider that $p_M'$ is functionally the same as $p_M$ but with added parameters arising from $R$. And so, by finding a retriever $R$ which maximizes the DPO metric, we are in effect fine-tuning the model $p_M'$.

\subsection{Retrieval Algorithm}

This section details the core algorithm employed for few-shot demonstration retrieval, which leverages a greedy strategy to maximize a combination of three key scores: query relevance, demonstration diversity, and demonstration quality bias.
The full retrieval algorithm is presented in Algorithm \ref{alg:mmr}.

\subsubsection{Query Relevance}
The relevance score between the query and each candidate demonstration is calculated using the cosine similarity of their respective \verb|BERT| embeddings \cite{devlinNACL2019}. By computing the cosine similarity between the query embedding and the embedding of each demonstration's query, we obtain a score that reflects the topical similarity and semantic alignment between the query and the candidate demonstration.

\subsubsection{Retrieved Demonstrations Diversity}
To promote diversity in the retrieved demonstrations and avoid redundancy, we incorporate the Maximal Marginal Relevance (MMR) algorithm. MMR iteratively selects the items that maximizes the combined score of relevance to the query and dissimilarity to the previously chosen items. This ensures a balance between retrieving relevant items and ensuring they cover a variety of information. A parameter, $\lambda_d$, is used to control the trade-off between relevance and diversity. Higher values of $\lambda_d$ prioritize relevance, whereas lower values prioritize diversity.

\subsubsection{Demonstration Quality Bias}
While the pre-trained \verb|BERT| embeddings capture semantic relationships, they do not inherently account for the quality of the few-shot demonstrations. To address this, we explicitly introduce a demonstration quality bias term related to the popularity of an item in a training dataset. This score is computed using the log perplexity of the demonstration answer $a$, given the demonstration question $q$.
$$
\frac{1}{|a|} \sum_{a_i \in a} \log p_M(a_i \mid q)
$$

This can be interpreted as measuring the probability of the correct answer $a$ given the query $q$, normalized to the length of the answer. This can also been interpreted as a proxy for a popularity bias, akin to the number of connections of an item in graph-based retrieval algorithms like recommender systems. Like in the article \cite{gonenEMNLP2023}, the intuition is that the more frequently a related sequence of tokens occurs in the pre-training dataset of the LLM, the more likely the model will be able to extract its relevant information.
Rather than directly analyzing the massive amount of text data (often trillions of tokens) used to pre-train the LLM, we focus on the perplexity of the sequence. Perplexity acts as a proxy, indicating how surprised the LLM is by the sequence, essentially, how well it aligns with what the LLM expects to see.
A parameter $\lambda_b$ controls the trade-off between relevance/diversity and quality bias. Lower values of $\lambda_b$ emphasize high-quality demonstrations.

\begin{algorithm}
\caption{MMR with quality bias}\label{alg:mmr}
\begin{algorithmic}
\Require $1 \leq k \leq n$;\; $0 \leq \lambda_d \leq 1$;\; $0 \leq \lambda_b \leq 1$
\Require $Q \in \mathbb{R}^d$ \Comment{query embedding}
\Require $\forall 1 \leq i \leq n$,\; $E_i \in \mathbb{R}^d$ \Comment{example question embedding}
\Require $\forall 1 \leq i \leq n$,\; $b_i \in \mathbb{R}$ \Comment{example quality bias}
\State $Q \gets \frac{Q}{\|Q\|}$;\; $\forall 1 \leq i \leq n$,\; $E_i \gets \frac{E_i}{\|E_i\|}$
\State $\forall 1 \leq i \leq n$,\; $v_i \gets \lambda_b Q E_i^\top + (1-\lambda_b) b_i$
\State $C_1 \gets \arg\max_i v_i$
\For{$2 \leq s \leq k$}
    \State $\forall 1 \leq i \leq n$,\; $m_i \gets \max_{1 \leq j < s} E_i E_{C_j}^\top$
    \State $\forall 1 \leq i \leq n$,\; $w_i \gets \lambda_d v_i - (1-\lambda_d) m_i$
    \State $C_s \gets \arg\max_{i \notin \{C_1, \dots, C_{s-1}\}} w_i$
\EndFor
\end{algorithmic}
\end{algorithm}

\section{Experiments}

\subsection{Experimental Setup}

\subsubsection{The Dataset Choice}
We are interested in a publicly available dataset which meets the following criteria: it should have enough statistical power so that we can resolve small differences in accuracy, ideally it will have hundreds of examples or more; it doesn't need to have vast amounts of data as this isn't a typical setting for few-shot learning, and the cost of conducting experiments can become burdensome; it should provide correct and incorrect answers so that we can report classification metrics from Section \ref{sec:classsification-metrics}; it should be big enough to contain similar examples with partially redundant information so the use of diversity can improve collective information presented to the LLM in a context.

\subsubsection{The TruthfulQA Dataset}
We chose to conduct our experiments using the TruthfulQA dataset \cite{Lin2021TruthfulQAMH} which meets these requirements. The dataset contains 817 distinct examples, which yields a standard error in the range of 1\% to 2\% for accuracy measures in the range of 90\% to 50\%. Each example contains a single query, and a variable number of correct and incorrect answers. And by considering each distinct $(q, a)$ pair as a demonstration for the purpose of building a context, the retriever is faced with similar demonstrations as multiple $(q, a)$ pairs share the same query (on average, the dataset contains 3.5 correct answers for each query).

\subsubsection{Generating Demonstrations Pairs and Triplets}
The dataset is used in three different ways in this paper:

\begin{itemize}
    \item $\mathcal{D}_{MC}$: this is the dataset as described in \cite{Lin2021TruthfulQAMH}. It contains 817 examples, each of which contains a variable number of correct and incorrect answers. The metrics MC1, MC2 and MC3, which can accept an arbitrary number of correct and incorrect answers as inputs, are calculated over this dataset.
    \item $\mathcal{D}_{DPO}$: this is the set of every distinct $(q, a, \bar{a})$  triple contained in $\mathcal{D}_{MC}$. It contains 12,485 such triplets. The DPO metric is calculated over this dataset.
    \item $\mathcal{D}_{ICL}$: this is the set of every distinct $(q, a)$ pairs contained in $\mathcal{D}_{MC}$. It contains 2,846 such pairs. This is the set of demonstrations from which a context is drawn. That is, $C \subset \mathcal{D}_{ICL}$.
\end{itemize}

When calculating a metric score for an example $(q_i, \mathbf{a}_i, \mathbf{\bar{a}}_i)$, all demonstrations with the query $q_i$ are left out from the demonstrations available for inclusion in the context. In this manner, the correct answers $\mathbf{a}_i$ are not included in the context when the LLM is presented with query $q_i$. 

\subsubsection{The Language Models}
We conducted our experiments using four noteworthy LLMs:
the smaller base text-completion model \textit{Mistral-7B-v0.1} (7B parameters), and the larger instruct-fine-tuned mixture of models \textit{Mixtral-8x22B-Instruct-v0.1} (141B parameters) from Mistral \cite{jiang2023arXiv};
as well as a smaller chat-tuned model \textit{Llama-3-8B-chat} (8B parameters) and a larger chat-tuned model \textit{Llama-3-70B-chat} (70B parameters) from Llama
\footnote{\url{https://llama.meta.com/llama3/}}.

All four models are open-weights LLMs, meaning their internal parameters are publicly available for scrutiny and potential fine-tuning.
These modern models stand out for achieving impressive performance on various tasks despite their relatively compact size. This efficiency makes it an attractive option for resource-constrained environments where deploying colossal models might not be feasible.

\subsection{Implementation Details and Ablation Study}
We implemented the Algorithm \ref{alg:mmr} and metrics from Section \ref{sec:classsification-metrics} in python. We computed \verb|BERT| embeddings using the package \textit{sentence\_transformers} \footnote{\url{https://sbert.net/}}, and implemented the retrieval algorithms in \textit{numpy}. We queried all LLM models using the Together API \footnote{\url{https://docs.together.ai/docs/inference-python}}.

We did not perform hyper-parameter tuning, and fixed the two parameters to $\lambda_d = 0.75$ and $\lambda_b = 0.95$ in all experiments.
We fixed the amount of retrieved demonstrations to $k=6$, matching the number of few-shot examples from the fixed primer example from the TruthfulQA paper \cite{Lin2021TruthfulQAMH}.

To measure the impact of the separated components of Algorithm \ref{alg:mmr}, relevance, diversity, and bias, we implemented variants of the retrieval algorithm using only one or two of the three components:

\begin{itemize}
\item Fix: fixed primer examples \cite{Lin2021TruthfulQAMH}
\item Bias: Pure quality bias \cite{gonenEMNLP2023}
\item Rel: Pure semantic similarity \cite{liuDEELIO2022} (\verb|KATE|)
\item Rel+Bias: Semantic similarity plus quality bias
\item Rel+Div: Semantic similarity plus diversity \cite{yeFINDINGSACL2023}
\item Rel+Bias+Div: Algorithm \ref{alg:mmr}
\end{itemize}

\subsection{Main Results}

We present the experimental metrics for the 6 retrievers for the 4 different LLMs:
in Table \ref{tab:res-mistral} and \ref{tab:res-mixtral-8x22B} for the Mistral models,
and Table \ref{tab:res-llama} and \ref{tab:res-llama-70b} for the Llama-3 models.

\begin{table*}[t]
\begin{minipage}{0.5\columnwidth}
  \caption{Evaluation Metrics with \textit{Mistral-7B-v0.1}}
  \label{tab:res-mistral}
  \begin{tabular}{lrrrr|}
    \toprule
    Method & DPO & MC1 & MC2 & MC3\\
    \midrule
    \small
    Fix & -20.40 & 0.2815 & 0.2086 & 0.4285\\
    Bias & -33.56 & 0.2411 & 0.1652 & 0.3596\\
    Rel & -12.71 & 0.4455 & 0.3664 & 0.5925\\
    Rel+Bias & -13.63 & 0.4602 & 0.3663 & 0.5969\\
    Rel+Div & \textbf{-12.37} & \textbf{0.5177} & \textbf{0.3930} & \textbf{0.6616}\\
    Rel+Div+Bias & -14.54 & 0.4676 & 0.3592 & 0.6255\\
  \bottomrule
\end{tabular}
\end{minipage}%
\begin{minipage}{0.5\columnwidth}
  \caption{Evaluation Metrics with \textit{Mixtral-8x22B-Instruct-v0.1}}
  \label{tab:res-mixtral-8x22B}
  \begin{tabular}{|lrrrr}
    \toprule
    Method & DPO & MC1 & MC2 & MC3\\
    \midrule
    Fix & -19.06 & 0.5202 & 0.3896 & 0.6799\\
    Bias & -27.30 & 0.4382 & 0.3096 & 0.5948\\
    Rel & -15.17 & 0.6193 & 0.5004 & 0.7616\\
    Rel+Bias & -14.77 & 0.6389 & 0.5080 & 0.7657\\
    Rel+Div & \textbf{-12.67} & \textbf{0.6879} & \textbf{0.5181} & \textbf{0.8092}\\
    Rel+Div+Bias & -13.29 & 0.6573 & 0.5071 & 0.7924\\
  \bottomrule
\end{tabular}
\end{minipage}
\end{table*}

\begin{table*}[t]
\begin{minipage}{0.5\columnwidth}
  \caption{Evaluation Metrics with \textit{Llama-3-8B-chat}}
  \label{tab:res-llama}
  \begin{tabular}{lrrrr|}
    \toprule
    Method & DPO & MC1 & MC2 & MC3\\
    \midrule
    Fix & -22.12 & 0.3623 & 0.2709 & 0.5195\\
    Bias & -17.55 & 0.3831 & 0.2876 & 0.5729\\
    Rel & -17.20 & 0.4920 & 0.4046 & 0.6518\\
    Rel+Bias & -17.14 & 0.5043 & 0.4083 & 0.6570\\
    Rel+Div & -16.14 & \textbf{0.5520} & \textbf{0.4173} & \textbf{0.7009}\\
    Rel+Div+Bias & \textbf{-15.80} & 0.5177 & 0.4007 & 0.6841 \\
  \bottomrule
\end{tabular}
\end{minipage}%
\begin{minipage}{0.5\columnwidth}
  \caption{Evaluation Metrics with \textit{Llama-3-70B-chat}}
  \label{tab:res-llama-70b}
  \begin{tabular}{|lrrrr}
    \toprule
    Method & DPO & MC1 & MC2 & MC3\\
    \midrule
    Fix & -23.05 & 0.4382 & 0.3375 & 0.6184\\
    Bias & -20.41 & 0.4455 & 0.3303 & 0.6424\\
    Rel & -19.02 & 0.5483 & 0.4482 & 0.6958\\
    Rel+Bias & -19.09 & 0.5532 & 0.4495 & 0.7054\\
    Rel+Div & -13.93 & \textbf{0.6389} & \textbf{0.4834} & \textbf{0.7758}\\
    Rel+Div+Bias & \textbf{-13.72} & 0.6022 & 0.4621 & 0.7583 \\
  \bottomrule
\end{tabular}
\end{minipage}
\end{table*}

Our evaluation relies on a combination of metrics to assess the effectiveness of different retrieval strategies for ICL. The normalized DPO metric provides the most valuable insights for each LLM individually but cannot be directly compared across models. The three additional classification metrics allow for objective performance comparisons across models. However, these metrics are susceptible to bias based on token sequence length.

The impact of few-shot learning is best seen by comparing the three MC metrics for \textit{Rel+Div} for a smaller model against \textit{Fix} for a larger model: the smaller models (7B and 8B parameters) enriched with ICL RAG are essentially matching or outperforming the bigger models (141B and 70B parameters) without ICL RAG.

The results consistently demonstrate that incorporating both relevance and diversity into the retrieval strategy leads to superior performance across all metrics and for both LLMs.
For all models, and for all metrics, \textit{Rel+Div} largely outperforms \textit{Rel}.
This finding reinforces the importance of not just retrieving relevant demonstrations but also ensuring a diverse set that maximizes the informative value for the LLM.

Interestingly, the impact of the low perplexity bias yields contrasting results. For both Mistral models, adding this bias results in a decline in performance on almost all metrics. Conversely, both Llama-3 models exhibit overall improvement with the low perplexity bias, in particular with the DPO metric. This intriguing observation suggests that LLM-dependent hyper-parameter tuning of $\lambda_b$ might be necessary to optimize retrieval strategies for specific models. Alternatively, the low perplexity bias itself may benefit from further refinement. Using an opposite intuition, we may argue that instead of prioritizing demonstrations the LLM already finds likely, introducing demonstrations that surprise the model the most could be beneficial for certain LLMs, potentially maximizing the learning impact of each demonstration. These findings open exciting new avenues for future research in ICL retrieval strategies, creating a parallel with novelty and serendipity concepts in recommender systems.

\subsection{Calibrating Diversity using DPO}

\begin{figure}[h]
  \centering
  \includegraphics[width=0.75 \linewidth]{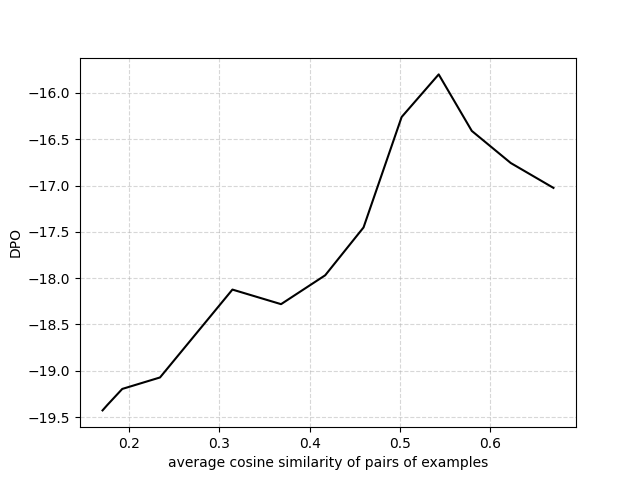}
  \caption{Diversity Metric and DPO. Non-monotonous relationship between a diversity metric, the average cosine similarity between embedding pairs, and the quality metric DPO. Obtained by varying $\lambda_d$ with \textit{Llama-3-8B-chat} and $k=6$.}
  \label{fig:div_vs_dpo}
\end{figure}

Calibrating the amount of diversity in the retrieved set is crucial when optimizing ICL retrieval. We highlight the difficulty of achieving this without our proposed methodology by demonstrating the non-monotonous relationship between the amount of diversity in the retrieved demonstrations and the resulting benefit to the LLM performance. To quantify diversity, we calculate the average cosine similarity between the \verb|BERT| embeddings of each demonstration pair within the retrieved set.  The LLM's benefit is measured using the DPO metric.  We then systematically vary $\lambda_d$ while keeping the LLM fixed (\textit{Llama-3-8B-chat}), the quality bias fixed ($\lambda_b=0.95$), and the number of retrieved demonstrations constant ($k=6$) to observe the empirical correlation between diversity and DPO. The results are visualized in Figure \ref{fig:div_vs_dpo}. This experiment underscores the importance of a metric measuring the impact of the retrieved context on the LLM, like DPO. Without such a metric, it would be challenging to effectively calibrate $\lambda_d$ and achieve the optimal balance between relevance and diversity in the retrieved demonstrations.

\section{Discussion: Real-World RAG Systems}

While the importance of diversity in ICL retrieval is paramount, we note that readily available RAG systems rarely implement it directly within the core retrieval algorithm. There are several practical considerations to keep in mind for successful deployment.

\subsection{Balancing Performance and Efficiency}
Retrieval latency is crucial at scale. Exhaustive, brute-force nearest neighbor search is computationally expensive and impractical. Instead, real-world systems leverage efficient indexing techniques and approximated \verb|kNN| algorithms, as described in \cite{beygelzimerICML2006}, to ensure fast retrieval times. This approach is essential for handling large datasets while maintaining responsiveness.
To seamlessly integrate with existing retrieval engines and leverage their optimized search capabilities, a retrieval algorithm for RAG must ensure its data is stored in a format compatible with these engines. Commonly indexed data structures include text itself or low-dimensional dense vector embeddings. By adhering to these indexing practices, RAG systems can effectively leverage the power of existing retrieval engines and achieve fast, scalable retrieval of informative examples.

\subsection{ICL RAG versus Fine-Tuning}
The computational cost of ICL may be evaluated against the cost of fine-tuning. For instance, consider a large LLM like \verb|gpt-3.5| with a current price 6x larger per input tokens between fine-tuned or default model
\footnote{In May 2024, the price of \textit{gpt-turbo-0125} is \$0.5/M input tokens, \$1.5/M output tokens; and fine-tuned price of \$3/M input tokens, \$6/M output tokens, \$8M fine-tuning tokens}.
While ICL requires additional input tokens, it is guaranteed to offer cost savings compared to fine-tuning when $k<6$ with this model.

An interesting contrast between ICL and fine-tuning is highlighted in the paper \cite{liuNEURIPS2022}. The paper argues that fine-tuning can be more efficient than few-shot ICL in terms of cost and latency due to the super-linear increase in LLM latency with growing prompt sizes. However, this latency concern is less relevant with inference-throughput optimized LLM systems built with large GPU clusters, such as commonly used APIs. In these systems, the observed latency remains independent of the prompt size. From a latency perspective, adding ICL demonstrations can be considered free. Additionally, the paper suggests that ICL requires scanning through 100\% of the demonstrations at query time. However this does not hold when employing real retrieval engines with indexing and approximate \verb|kNN|, which significantly reduce the number of examples scanned during retrieval.

Furthermore, building a curated database of few-shot demonstrations offers significant advantages to practitioners. These demonstrations are not specific to a single LLM but can be readily utilized with any LLM architecture. This eliminates vendor lock-in and lets practitioners leverage the best LLM for the task at hand without concerns about compatibility. Perhaps even more importantly, a well-maintained database of few-shot examples automatically benefits from the continuous advancements in LLM technology. As newer, more powerful pre-trained LLMs become rapidly available, existing demonstrations can be used to enrich them quickly. This ensures applications leverage the latest capabilities without the need to completely re-engineer workflows. This reusability and adaptability position our few-shot learning engine as a powerful tool for harnessing the ever-evolving potential of LLMs to solve real business challenges.

\subsection{Achieving State-of-the-Art Retrieval with Available Tools}
Traditional full-text search algorithms like BM25 lead to empirically lower ICL quality. Vector stores offer a more suitable solution for efficient retrieval based on semantic similarity.  Numerous vendors provide vector store solutions, and they can be broadly categorized as follows:

In-Memory vector indexes, such as \verb|FAISS| and \verb|nmslib|, offer exceptional speed with minimal setup complexity, but limited scalability for larger datasets. They may not implement in-place addition or deletion of the indexed vectors.
Self-Hosted vector databases, such as \verb|Elasticsearch| and \verb|Postgres|, provide a balance between scalability and performance, at a much larger setup complexity. They typically implement efficient addition and deletion of the indexed vectors.
SaaS vector stores, such as \verb|Pinecone| and \verb|VertexAI|, offer a convenient option with pre-configured infrastructure and almost no setup complexity.
We invite the reader to consult the lists of integrated vector stores
of LangChain \footnote{\url{https://python.langchain.com/docs/integrations/vectorstores/}}
and LlamaIndex \footnote{\url{https://docs.llamaindex.ai/en/stable/module_guides/storing/vector_stores/}}
for near-exhaustive lists of available tools.

Due to the complexities of incorporating such rules directly within retrieval indexing algorithm \cite{liVLDB2013}, none of the solutions known to the authors from any of the above category implements diversity or quality biasing of the result.
A common heuristic to mitigate this problem is to retrieve a larger set of candidate examples (e.g., double the desired number) and then apply diversity techniques like MMR on the retrieved candidates as a post-processing step.
Quality biasing can be indirectly achieved by modifying the indexed embeddings themselves. For instance, reducing the norm of embeddings associated with low-quality content can nudge the retrieval algorithm towards higher-quality examples. An exact implementation in the context of cosine-similarity or dot-product relevance is to add an additional column storing the quality bias, and set the corresponding value to $1$ in the embedding of the query.

While vector search offers a powerful foundation for practical ICL retrieval, it often lacks native support for essential considerations like diversity or quality bias. These aspects are crucial for ensuring informative and effective retrieval of few-shot learning examples. Existing tools for recommendation engines, on the other hand, often excel in these areas. Recommendation engines natively incorporate rules that promote diversity by recommending a variety of items, or quality bias by prioritizing most popular products. Future research directions as well as practical systems for ICL retrieval could explore adapting or integrating these well-established techniques from recommender systems to further enhance the effectiveness and sophistication of few-shot learning through information retrieval. State-of-the-art ICL for real-world applications can be achieved by combining the strengths of vector search with the established "diversity-aware" retrieval approaches from recommender systems.

\section{Conclusion}

This paper explored the critical role of information retrieval in ICL for few-shot learning with Large Language Models. Our work identified key desirable properties for ICL retrieval systems. We demonstrated that state-of-the-art retrieval in this domain resembles recommender systems under the item cold-start problems. Unlike traditional information retrieval prioritizing for exact recall, our approach emphasizes discovery by maximizing the collective information gain from retrieved demonstrations. This necessitates balancing query relevance, quality scoring, and diversity to ensure a variety of informative examples are surfaced.
Furthermore, we propose a novel evaluation method for ICL retrieval based on the subsequent performance of the enriched LLM on NLP tasks. This approach eliminates the need for subjective diversity scores, a challenge in information retrieval evaluation.
Our findings demonstrate the significant impact of diversity and quality bias in retrieving demonstrations for ICL. By incorporating these well-established techniques from recommender systems, we can unlock the full potential of ICL for few-shot learning and empower LLMs to tackle real-world tasks with limited data.

\section{Acknowledgments}

To Ching-Wei Chen, for finding the name \textit{RAGSys}, and the entire Crossing Minds team for their support during this research.

\bibliography{ir_rag_sigir24}

\begin{thebibliography}{52}
\expandafter\ifx\csname natexlab\endcsname\relax\def\natexlab#1{#1}\fi
\providecommand{\url}[1]{\texttt{#1}}
\providecommand{\href}[2]{#2}
\providecommand{\path}[1]{#1}
\providecommand{\DOIprefix}{doi:}
\providecommand{\ArXivprefix}{arXiv:}
\providecommand{\URLprefix}{URL: }
\providecommand{\Pubmedprefix}{pmid:}
\providecommand{\doi}[1]{\href{http://dx.doi.org/#1}{\path{#1}}}
\providecommand{\Pubmed}[1]{\href{pmid:#1}{\path{#1}}}
\providecommand{\bibinfo}[2]{#2}
\ifx\xfnm\relax \def\xfnm[#1]{\unskip,\space#1}\fi
\bibitem[{Brown et~al.(2020)Brown, Mann, Ryder, Subbiah, Kaplan, Dhariwal, Neelakantan, Shyam, Sastry, Askell et~al.}]{brownNEURIPS2020}
\bibinfo{author}{T.~Brown}, \bibinfo{author}{B.~Mann}, \bibinfo{author}{N.~Ryder}, \bibinfo{author}{M.~Subbiah}, \bibinfo{author}{J.~D. Kaplan}, \bibinfo{author}{P.~Dhariwal}, \bibinfo{author}{A.~Neelakantan}, \bibinfo{author}{P.~Shyam}, \bibinfo{author}{G.~Sastry}, \bibinfo{author}{A.~Askell}, et~al.,
\newblock \bibinfo{title}{Language models are few-shot learners},
\newblock \bibinfo{journal}{Advances in neural information processing systems} \bibinfo{volume}{33} (\bibinfo{year}{2020}) \bibinfo{pages}{1877--1901}.
\bibitem[{Kojima et~al.(2022)Kojima, Gu, Reid, Matsuo, and Iwasawa}]{kojimaNEURIPS2022}
\bibinfo{author}{T.~Kojima}, \bibinfo{author}{S.~S. Gu}, \bibinfo{author}{M.~Reid}, \bibinfo{author}{Y.~Matsuo}, \bibinfo{author}{Y.~Iwasawa},
\newblock \bibinfo{title}{Large language models are zero-shot reasoners},
\newblock in: \bibinfo{editor}{S.~Koyejo}, \bibinfo{editor}{S.~Mohamed}, \bibinfo{editor}{A.~Agarwal}, \bibinfo{editor}{D.~Belgrave}, \bibinfo{editor}{K.~Cho}, \bibinfo{editor}{A.~Oh} (Eds.), \bibinfo{booktitle}{Advances in Neural Information Processing Systems}, volume~\bibinfo{volume}{35}, \bibinfo{publisher}{Curran Associates, Inc.}, \bibinfo{year}{2022}, pp. \bibinfo{pages}{22199--22213}. \URLprefix \url{https://proceedings.neurips.cc/paper_files/paper/2022/file/8bb0d291acd4acf06ef112099c16f326-Paper-Conference.pdf}.
\bibitem[{Hu et~al.(2022)Hu, yelong shen, Wallis, Allen-Zhu, Li, Wang, Wang, and Chen}]{HuICLR2022}
\bibinfo{author}{E.~J. Hu}, \bibinfo{author}{yelong shen}, \bibinfo{author}{P.~Wallis}, \bibinfo{author}{Z.~Allen-Zhu}, \bibinfo{author}{Y.~Li}, \bibinfo{author}{S.~Wang}, \bibinfo{author}{L.~Wang}, \bibinfo{author}{W.~Chen},
\newblock \bibinfo{title}{Lo{RA}: Low-rank adaptation of large language models},
\newblock in: \bibinfo{booktitle}{International Conference on Learning Representations}, \bibinfo{year}{2022}. \URLprefix \url{https://openreview.net/forum?id=nZeVKeeFYf9}.
\bibitem[{Luo et~al.(2023)Luo, Yang, Meng, Li, Zhou, and Zhang}]{Luo2023AnES}
\bibinfo{author}{Y.~Luo}, \bibinfo{author}{Z.~Yang}, \bibinfo{author}{F.~Meng}, \bibinfo{author}{Y.~Li}, \bibinfo{author}{J.~Zhou}, \bibinfo{author}{Y.~Zhang},
\newblock \bibinfo{title}{An empirical study of catastrophic forgetting in large language models during continual fine-tuning},
\newblock \bibinfo{journal}{ArXiv} \bibinfo{volume}{abs/2308.08747} (\bibinfo{year}{2023}). \URLprefix \url{https://api.semanticscholar.org/CorpusID:261031244}.
\bibitem[{Berglund et~al.(2023)Berglund, Tong, Kaufmann, Balesni, Stickland, Korbak, and Evans}]{Berglund2023TheRC}
\bibinfo{author}{L.~Berglund}, \bibinfo{author}{M.~Tong}, \bibinfo{author}{M.~Kaufmann}, \bibinfo{author}{M.~Balesni}, \bibinfo{author}{A.~C. Stickland}, \bibinfo{author}{T.~Korbak}, \bibinfo{author}{O.~Evans},
\newblock \bibinfo{title}{The reversal curse: Llms trained on "a is b" fail to learn "b is a"},
\newblock \bibinfo{journal}{ArXiv} \bibinfo{volume}{abs/2309.12288} (\bibinfo{year}{2023}). \URLprefix \url{https://api.semanticscholar.org/CorpusID:262083829}.
\bibitem[{Bai et~al.(2023)Bai, Chen, Wang, Xiong, and Mei}]{BaiNEURIPS2023}
\bibinfo{author}{Y.~Bai}, \bibinfo{author}{F.~Chen}, \bibinfo{author}{H.~Wang}, \bibinfo{author}{C.~Xiong}, \bibinfo{author}{S.~Mei},
\newblock \bibinfo{title}{Transformers as statisticians: Provable in-context learning with in-context algorithm selection},
\newblock in: \bibinfo{editor}{A.~Oh}, \bibinfo{editor}{T.~Neumann}, \bibinfo{editor}{A.~Globerson}, \bibinfo{editor}{K.~Saenko}, \bibinfo{editor}{M.~Hardt}, \bibinfo{editor}{S.~Levine} (Eds.), \bibinfo{booktitle}{Advances in Neural Information Processing Systems}, volume~\bibinfo{volume}{36}, \bibinfo{publisher}{Curran Associates, Inc.}, \bibinfo{year}{2023}, pp. \bibinfo{pages}{57125--57211}. \URLprefix \url{https://proceedings.neurips.cc/paper_files/paper/2023/file/b2e63e36c57e153b9015fece2352a9f9-Paper-Conference.pdf}.
\bibitem[{Ram et~al.(2023)Ram, Levine, Dalmedigos, Muhlgay, Shashua, Leyton-Brown, and Shoham}]{RamTACL2023}
\bibinfo{author}{O.~Ram}, \bibinfo{author}{Y.~Levine}, \bibinfo{author}{I.~Dalmedigos}, \bibinfo{author}{D.~Muhlgay}, \bibinfo{author}{A.~Shashua}, \bibinfo{author}{K.~Leyton-Brown}, \bibinfo{author}{Y.~Shoham},
\newblock \bibinfo{title}{{In-Context Retrieval-Augmented Language Models}},
\newblock \bibinfo{journal}{Transactions of the Association for Computational Linguistics} \bibinfo{volume}{11} (\bibinfo{year}{2023}) \bibinfo{pages}{1316--1331}. \DOIprefix\doi{10.1162/tacl_a_00605}.
\bibitem[{Dong et~al.(2023)Dong, Li, Dai, Zheng, Wu, Chang, Sun, Xu, Li, and Sui}]{QingxiuARXIV2023}
\bibinfo{author}{Q.~Dong}, \bibinfo{author}{L.~Li}, \bibinfo{author}{D.~Dai}, \bibinfo{author}{C.~Zheng}, \bibinfo{author}{Z.~Wu}, \bibinfo{author}{B.~Chang}, \bibinfo{author}{X.~Sun}, \bibinfo{author}{J.~Xu}, \bibinfo{author}{L.~Li}, \bibinfo{author}{Z.~Sui},
\newblock \bibinfo{title}{A survey for in-context learning},
\newblock \bibinfo{journal}{CoRR} \bibinfo{volume}{abs/2301.00234} (\bibinfo{year}{2023}).
\bibitem[{{Machlab} and {Battle}(2024)}]{MachlabARXIV2024}
\bibinfo{author}{D.~{Machlab}}, \bibinfo{author}{R.~{Battle}},
\newblock \bibinfo{title}{{LLM In-Context Recall is Prompt Dependent}},
\newblock \bibinfo{journal}{CoRR}  (\bibinfo{year}{2024}) \bibinfo{pages}{arXiv:2404.08865}. \DOIprefix\doi{10.48550/arXiv.2404.08865}. \href{http://arxiv.org/abs/2404.08865}{{\tt arXiv:2404.08865}}.
\bibitem[{Clarke et~al.(2008)Clarke, Kolla, Cormack, Vechtomova, Ashkan, B{\"u}ttcher, and MacKinnon}]{clarkeSIGIR2008}
\bibinfo{author}{C.~L. Clarke}, \bibinfo{author}{M.~Kolla}, \bibinfo{author}{G.~V. Cormack}, \bibinfo{author}{O.~Vechtomova}, \bibinfo{author}{A.~Ashkan}, \bibinfo{author}{S.~B{\"u}ttcher}, \bibinfo{author}{I.~MacKinnon},
\newblock \bibinfo{title}{Novelty and diversity in information retrieval evaluation},
\newblock in: \bibinfo{booktitle}{Proceedings of the 31st Annual International ACM SIGIR Conference on Research and Development in Information Retrieval}, SIGIR '08, \bibinfo{publisher}{Association for Computing Machinery}, \bibinfo{address}{New York, NY, USA}, \bibinfo{year}{2008}, p. \bibinfo{pages}{659–666}. \DOIprefix\doi{10.1145/1390334.1390446}.
\bibitem[{Yu et~al.(2023)Yu, Iter, Wang, Xu, Ju, Sanyal, Zhu, Zeng, and Jiang}]{yuICLR2023}
\bibinfo{author}{W.~Yu}, \bibinfo{author}{D.~Iter}, \bibinfo{author}{S.~Wang}, \bibinfo{author}{Y.~Xu}, \bibinfo{author}{M.~Ju}, \bibinfo{author}{S.~Sanyal}, \bibinfo{author}{C.~Zhu}, \bibinfo{author}{M.~Zeng}, \bibinfo{author}{M.~Jiang},
\newblock \bibinfo{title}{Generate rather than retrieve: Large language models are strong context generators},
\newblock in: \bibinfo{booktitle}{International Conference for Learning Representation (ICLR)}, \bibinfo{year}{2023}.
\bibitem[{Gonen et~al.(2023)Gonen, Iyer, Blevins, Smith, and Zettlemoyer}]{gonenEMNLP2023}
\bibinfo{author}{H.~Gonen}, \bibinfo{author}{S.~Iyer}, \bibinfo{author}{T.~Blevins}, \bibinfo{author}{N.~Smith}, \bibinfo{author}{L.~Zettlemoyer},
\newblock \bibinfo{title}{Demystifying prompts in language models via perplexity estimation},
\newblock in: \bibinfo{editor}{H.~Bouamor}, \bibinfo{editor}{J.~Pino}, \bibinfo{editor}{K.~Bali} (Eds.), \bibinfo{booktitle}{Findings of the Association for Computational Linguistics: EMNLP 2023}, \bibinfo{publisher}{Association for Computational Linguistics}, \bibinfo{address}{Singapore}, \bibinfo{year}{2023}, pp. \bibinfo{pages}{10136--10148}. \DOIprefix\doi{10.18653/v1/2023.findings-emnlp.679}.
\bibitem[{Agrawal et~al.(2023)Agrawal, Zhou, Lewis, Zettlemoyer, and Ghazvininejad}]{agrawalFINDINGSACL2023}
\bibinfo{author}{S.~Agrawal}, \bibinfo{author}{C.~Zhou}, \bibinfo{author}{M.~Lewis}, \bibinfo{author}{L.~Zettlemoyer}, \bibinfo{author}{M.~Ghazvininejad},
\newblock \bibinfo{title}{In-context examples selection for machine translation},
\newblock in: \bibinfo{editor}{A.~Rogers}, \bibinfo{editor}{J.~Boyd-Graber}, \bibinfo{editor}{N.~Okazaki} (Eds.), \bibinfo{booktitle}{Findings of the Association for Computational Linguistics: ACL 2023}, \bibinfo{publisher}{Association for Computational Linguistics}, \bibinfo{address}{Toronto, Canada}, \bibinfo{year}{2023}, pp. \bibinfo{pages}{8857--8873}. \DOIprefix\doi{10.18653/v1/2023.findings-acl.564}.
\bibitem[{Robertson et~al.(1995)Robertson, Walker, Jones, Hancock-Beaulieu, and Gatford}]{robertsonNIST1995}
\bibinfo{author}{S.~Robertson}, \bibinfo{author}{S.~Walker}, \bibinfo{author}{S.~Jones}, \bibinfo{author}{M.~M. Hancock-Beaulieu}, \bibinfo{author}{M.~Gatford},
\newblock \bibinfo{title}{Okapi at trec-3},
\newblock in: \bibinfo{booktitle}{Overview of the Third Text REtrieval Conference (TREC-3)}, \bibinfo{publisher}{Gaithersburg, MD: NIST}, \bibinfo{year}{1995}, pp. \bibinfo{pages}{109--126}. \URLprefix \url{https://www.microsoft.com/en-us/research/publication/okapi-at-trec-3/}.
\bibitem[{Rubin et~al.(2022)Rubin, Herzig, and Berant}]{rubinNACL2022}
\bibinfo{author}{O.~Rubin}, \bibinfo{author}{J.~Herzig}, \bibinfo{author}{J.~Berant},
\newblock \bibinfo{title}{Learning to retrieve prompts for in-context learning},
\newblock in: \bibinfo{editor}{M.~Carpuat}, \bibinfo{editor}{M.-C. de~Marneffe}, \bibinfo{editor}{I.~V. Meza~Ruiz} (Eds.), \bibinfo{booktitle}{Proceedings of the 2022 Conference of the North American Chapter of the Association for Computational Linguistics: Human Language Technologies}, \bibinfo{publisher}{Association for Computational Linguistics}, \bibinfo{address}{Seattle, United States}, \bibinfo{year}{2022}, pp. \bibinfo{pages}{2655--2671}. \DOIprefix\doi{10.18653/v1/2022.naacl-main.191}.
\bibitem[{{Wang} et~al.(2022){Wang}, {Yang}, {Huang}, {Jiao}, {Yang}, {Jiang}, {Majumder}, and {Wei}}]{wangARXIV2022}
\bibinfo{author}{L.~{Wang}}, \bibinfo{author}{N.~{Yang}}, \bibinfo{author}{X.~{Huang}}, \bibinfo{author}{B.~{Jiao}}, \bibinfo{author}{L.~{Yang}}, \bibinfo{author}{D.~{Jiang}}, \bibinfo{author}{R.~{Majumder}}, \bibinfo{author}{F.~{Wei}},
\newblock \bibinfo{title}{{Text Embeddings by Weakly-Supervised Contrastive Pre-training}},
\newblock \bibinfo{journal}{CoRR}  (\bibinfo{year}{2022}) \bibinfo{pages}{arXiv:2212.03533}. \DOIprefix\doi{10.48550/arXiv.2212.03533}. \href{http://arxiv.org/abs/2212.03533}{{\tt arXiv:2212.03533}}.
\bibitem[{Li et~al.(2023)Li, Lv, Yan, Lin, Zhu, Ni, Xie, Wang, and Qiu}]{liACL2023}
\bibinfo{author}{X.~Li}, \bibinfo{author}{K.~Lv}, \bibinfo{author}{H.~Yan}, \bibinfo{author}{T.~Lin}, \bibinfo{author}{W.~Zhu}, \bibinfo{author}{Y.~Ni}, \bibinfo{author}{G.~Xie}, \bibinfo{author}{X.~Wang}, \bibinfo{author}{X.~Qiu},
\newblock \bibinfo{title}{Unified demonstration retriever for in-context learning},
\newblock in: \bibinfo{editor}{A.~Rogers}, \bibinfo{editor}{J.~Boyd-Graber}, \bibinfo{editor}{N.~Okazaki} (Eds.), \bibinfo{booktitle}{Proceedings of the 61st Annual Meeting of the Association for Computational Linguistics (Volume 1: Long Papers)}, \bibinfo{publisher}{Association for Computational Linguistics}, \bibinfo{address}{Toronto, Canada}, \bibinfo{year}{2023}, pp. \bibinfo{pages}{4644--4668}. \DOIprefix\doi{10.18653/v1/2023.acl-long.256}.
\bibitem[{Ye et~al.(2023)Ye, Wu, Feng, Yu, and Kong}]{yeICML2023}
\bibinfo{author}{J.~Ye}, \bibinfo{author}{Z.~Wu}, \bibinfo{author}{J.~Feng}, \bibinfo{author}{T.~Yu}, \bibinfo{author}{L.~Kong},
\newblock \bibinfo{title}{Compositional exemplars for in-context learning},
\newblock in: \bibinfo{editor}{A.~Krause}, \bibinfo{editor}{E.~Brunskill}, \bibinfo{editor}{K.~Cho}, \bibinfo{editor}{B.~Engelhardt}, \bibinfo{editor}{S.~Sabato}, \bibinfo{editor}{J.~Scarlett} (Eds.), \bibinfo{booktitle}{Proceedings of the 40th International Conference on Machine Learning}, volume \bibinfo{volume}{202} of \textit{\bibinfo{series}{Proceedings of Machine Learning Research}}, \bibinfo{publisher}{PMLR}, \bibinfo{year}{2023}, pp. \bibinfo{pages}{39818--39833}. \URLprefix \url{https://proceedings.mlr.press/v202/ye23c.html}.
\bibitem[{Wang et~al.(2024)Wang, Yang, and Wei}]{wangEACL2024}
\bibinfo{author}{L.~Wang}, \bibinfo{author}{N.~Yang}, \bibinfo{author}{F.~Wei},
\newblock \bibinfo{title}{Learning to retrieve in-context examples for large language models},
\newblock in: \bibinfo{editor}{Y.~Graham}, \bibinfo{editor}{M.~Purver} (Eds.), \bibinfo{booktitle}{Proceedings of the 18th Conference of the European Chapter of the Association for Computational Linguistics (Volume 1: Long Papers)}, \bibinfo{publisher}{Association for Computational Linguistics}, \bibinfo{address}{St. Julian{'}s, Malta}, \bibinfo{year}{2024}, pp. \bibinfo{pages}{1752--1767}.
\bibitem[{Liu et~al.(2022)Liu, Shen, Zhang, Dolan, Carin, and Chen}]{liuDEELIO2022}
\bibinfo{author}{J.~Liu}, \bibinfo{author}{D.~Shen}, \bibinfo{author}{Y.~Zhang}, \bibinfo{author}{B.~Dolan}, \bibinfo{author}{L.~Carin}, \bibinfo{author}{W.~Chen},
\newblock \bibinfo{title}{What makes good in-context examples for {GPT}-3?},
\newblock in: \bibinfo{editor}{E.~Agirre}, \bibinfo{editor}{M.~Apidianaki}, \bibinfo{editor}{I.~Vuli{\'c}} (Eds.), \bibinfo{booktitle}{Proceedings of Deep Learning Inside Out (DeeLIO 2022): The 3rd Workshop on Knowledge Extraction and Integration for Deep Learning Architectures}, \bibinfo{publisher}{Association for Computational Linguistics}, \bibinfo{address}{Dublin, Ireland and Online}, \bibinfo{year}{2022}, pp. \bibinfo{pages}{100--114}. \DOIprefix\doi{10.18653/v1/2022.deelio-1.10}.
\bibitem[{Devlin et~al.(2019)Devlin, Chang, Lee, and Toutanova}]{devlinNACL2019}
\bibinfo{author}{J.~Devlin}, \bibinfo{author}{M.-W. Chang}, \bibinfo{author}{K.~Lee}, \bibinfo{author}{K.~Toutanova},
\newblock \bibinfo{title}{{BERT}: Pre-training of deep bidirectional transformers for language understanding},
\newblock in: \bibinfo{editor}{J.~Burstein}, \bibinfo{editor}{C.~Doran}, \bibinfo{editor}{T.~Solorio} (Eds.), \bibinfo{booktitle}{Proceedings of the 2019 Conference of the North {A}merican Chapter of the Association for Computational Linguistics: Human Language Technologies, Volume 1 (Long and Short Papers)}, \bibinfo{publisher}{Association for Computational Linguistics}, \bibinfo{address}{Minneapolis, Minnesota}, \bibinfo{year}{2019}, pp. \bibinfo{pages}{4171--4186}. \DOIprefix\doi{10.18653/v1/N19-1423}.
\bibitem[{Khattab and Zaharia(2020)}]{khattabSIGIR2020}
\bibinfo{author}{O.~Khattab}, \bibinfo{author}{M.~Zaharia},
\newblock \bibinfo{title}{Colbert: Efficient and effective passage search via contextualized late interaction over bert},
\newblock in: \bibinfo{booktitle}{Proceedings of the 43rd International ACM SIGIR Conference on Research and Development in Information Retrieval}, SIGIR '20, \bibinfo{publisher}{Association for Computing Machinery}, \bibinfo{address}{New York, NY, USA}, \bibinfo{year}{2020}, p. \bibinfo{pages}{39–48}. \DOIprefix\doi{10.1145/3397271.3401075}.
\bibitem[{Beygelzimer et~al.(2006)Beygelzimer, Kakade, and Langford}]{beygelzimerICML2006}
\bibinfo{author}{A.~Beygelzimer}, \bibinfo{author}{S.~Kakade}, \bibinfo{author}{J.~Langford},
\newblock \bibinfo{title}{Cover trees for nearest neighbor},
\newblock in: \bibinfo{booktitle}{Proceedings of the 23rd International Conference on Machine Learning}, ICML '06, \bibinfo{publisher}{Association for Computing Machinery}, \bibinfo{address}{New York, NY, USA}, \bibinfo{year}{2006}, p. \bibinfo{pages}{97–104}. \DOIprefix\doi{10.1145/1143844.1143857}.
\bibitem[{Levy et~al.(2023)Levy, Bogin, and Berant}]{levyACL2023}
\bibinfo{author}{I.~Levy}, \bibinfo{author}{B.~Bogin}, \bibinfo{author}{J.~Berant},
\newblock \bibinfo{title}{Diverse demonstrations improve in-context compositional generalization},
\newblock in: \bibinfo{editor}{A.~Rogers}, \bibinfo{editor}{J.~Boyd-Graber}, \bibinfo{editor}{N.~Okazaki} (Eds.), \bibinfo{booktitle}{Proceedings of the 61st Annual Meeting of the Association for Computational Linguistics (Volume 1: Long Papers)}, \bibinfo{publisher}{Association for Computational Linguistics}, \bibinfo{address}{Toronto, Canada}, \bibinfo{year}{2023}, pp. \bibinfo{pages}{1401--1422}. \DOIprefix\doi{10.18653/v1/2023.acl-long.78}.
\bibitem[{Su et~al.(2023)Su, Kasai, Wu, Shi, Wang, Xin, 0037, Ostendorf, Zettlemoyer, Smith, and 0009}]{suICLR2023}
\bibinfo{author}{H.~Su}, \bibinfo{author}{J.~Kasai}, \bibinfo{author}{C.~H. Wu}, \bibinfo{author}{W.~Shi}, \bibinfo{author}{T.~Wang}, \bibinfo{author}{J.~Xin}, \bibinfo{author}{R.~Z. 0037}, \bibinfo{author}{M.~Ostendorf}, \bibinfo{author}{L.~Zettlemoyer}, \bibinfo{author}{N.~A. Smith}, \bibinfo{author}{T.~Y. 0009},
\newblock \bibinfo{title}{Selective annotation makes language models better few-shot learners},
\newblock in: \bibinfo{booktitle}{The Eleventh International Conference on Learning Representations, ICLR 2023, Kigali, Rwanda, May 1-5, 2023}, \bibinfo{publisher}{OpenReview.net}, \bibinfo{year}{2023}. \URLprefix \url{https://openreview.net/pdf?id=qY1hlv7gwg}.
\bibitem[{Ye et~al.(2023)Ye, Iyer, Celikyilmaz, Stoyanov, Durrett, and Pasunuru}]{yeFINDINGSACL2023}
\bibinfo{author}{X.~Ye}, \bibinfo{author}{S.~Iyer}, \bibinfo{author}{A.~Celikyilmaz}, \bibinfo{author}{V.~Stoyanov}, \bibinfo{author}{G.~Durrett}, \bibinfo{author}{R.~Pasunuru},
\newblock \bibinfo{title}{Complementary explanations for effective in-context learning},
\newblock in: \bibinfo{editor}{A.~Rogers}, \bibinfo{editor}{J.~Boyd-Graber}, \bibinfo{editor}{N.~Okazaki} (Eds.), \bibinfo{booktitle}{Findings of the Association for Computational Linguistics: ACL 2023}, \bibinfo{publisher}{Association for Computational Linguistics}, \bibinfo{address}{Toronto, Canada}, \bibinfo{year}{2023}, pp. \bibinfo{pages}{4469--4484}. \DOIprefix\doi{10.18653/v1/2023.findings-acl.273}.
\bibitem[{Carbonell and Goldstein(1998)}]{carbonellSIGIR1998}
\bibinfo{author}{J.~Carbonell}, \bibinfo{author}{J.~Goldstein},
\newblock \bibinfo{title}{The use of mmr, diversity-based reranking for reordering documents and producing summaries},
\newblock in: \bibinfo{booktitle}{Proceedings of the 21st Annual International ACM SIGIR Conference on Research and Development in Information Retrieval}, SIGIR '98, \bibinfo{publisher}{Association for Computing Machinery}, \bibinfo{address}{New York, NY, USA}, \bibinfo{year}{1998}, p. \bibinfo{pages}{335–336}. \DOIprefix\doi{10.1145/290941.291025}.
\bibitem[{Zhai et~al.(2003)Zhai, Cohen, and Lafferty}]{zhaiSIGIR2003}
\bibinfo{author}{C.~X. Zhai}, \bibinfo{author}{W.~W. Cohen}, \bibinfo{author}{J.~Lafferty},
\newblock \bibinfo{title}{Beyond independent relevance: methods and evaluation metrics for subtopic retrieval},
\newblock in: \bibinfo{booktitle}{Proceedings of the 26th Annual International ACM SIGIR Conference on Research and Development in Informaion Retrieval}, SIGIR '03, \bibinfo{publisher}{Association for Computing Machinery}, \bibinfo{address}{New York, NY, USA}, \bibinfo{year}{2003}, p. \bibinfo{pages}{10–17}. \DOIprefix\doi{10.1145/860435.860440}.
\bibitem[{Agrawal et~al.(2009)Agrawal, Gollapudi, Halverson, and Ieong}]{agrawalWSDM2009}
\bibinfo{author}{R.~Agrawal}, \bibinfo{author}{S.~Gollapudi}, \bibinfo{author}{A.~Halverson}, \bibinfo{author}{S.~Ieong},
\newblock \bibinfo{title}{Diversifying search results},
\newblock in: \bibinfo{booktitle}{Proceedings of the Second ACM International Conference on Web Search and Data Mining}, WSDM '09, \bibinfo{publisher}{Association for Computing Machinery}, \bibinfo{address}{New York, NY, USA}, \bibinfo{year}{2009}, p. \bibinfo{pages}{5–14}. \URLprefix \url{https://doi.org/10.1145/1498759.1498766}. \DOIprefix\doi{10.1145/1498759.1498766}.
\bibitem[{Gollapudi and Sharma(2009)}]{gollapudiWWW2009}
\bibinfo{author}{S.~Gollapudi}, \bibinfo{author}{A.~Sharma},
\newblock \bibinfo{title}{An axiomatic approach for result diversification},
\newblock in: \bibinfo{booktitle}{Proceedings of the 18th International Conference on World Wide Web}, WWW '09, \bibinfo{publisher}{Association for Computing Machinery}, \bibinfo{address}{New York, NY, USA}, \bibinfo{year}{2009}, p. \bibinfo{pages}{381–390}. \DOIprefix\doi{10.1145/1526709.1526761}.
\bibitem[{Vieira et~al.(2011)Vieira, Razente, Barioni, Hadjieleftheriou, Srivastava, Traina, and Tsotras}]{vieiraICDE2011}
\bibinfo{author}{M.~R. Vieira}, \bibinfo{author}{H.~L. Razente}, \bibinfo{author}{M.~C.~N. Barioni}, \bibinfo{author}{M.~Hadjieleftheriou}, \bibinfo{author}{D.~Srivastava}, \bibinfo{author}{C.~Traina}, \bibinfo{author}{V.~J. Tsotras},
\newblock \bibinfo{title}{On query result diversification},
\newblock in: \bibinfo{booktitle}{Proceedings of the 2011 IEEE 27th International Conference on Data Engineering}, ICDE '11, \bibinfo{publisher}{IEEE Computer Society}, \bibinfo{address}{USA}, \bibinfo{year}{2011}, p. \bibinfo{pages}{1163–1174}. \DOIprefix\doi{10.1109/ICDE.2011.5767846}.
\bibitem[{Clarke et~al.(2009)Clarke, Craswell, and Soboroff}]{clarkeTREC2009}
\bibinfo{author}{C.~L. Clarke}, \bibinfo{author}{N.~Craswell}, \bibinfo{author}{I.~Soboroff},
\newblock \bibinfo{title}{Overview of the trec 2009 web track.},
\newblock in: \bibinfo{booktitle}{Trec}, volume~\bibinfo{volume}{9}, \bibinfo{year}{2009}, pp. \bibinfo{pages}{20--29}.
\bibitem[{Chapelle et~al.(2009)Chapelle, Metlzer, Zhang, and Grinspan}]{chapelleCIKM2009}
\bibinfo{author}{O.~Chapelle}, \bibinfo{author}{D.~Metlzer}, \bibinfo{author}{Y.~Zhang}, \bibinfo{author}{P.~Grinspan},
\newblock \bibinfo{title}{Expected reciprocal rank for graded relevance},
\newblock in: \bibinfo{booktitle}{Proceedings of the 18th ACM Conference on Information and Knowledge Management}, CIKM '09, \bibinfo{publisher}{Association for Computing Machinery}, \bibinfo{address}{New York, NY, USA}, \bibinfo{year}{2009}, p. \bibinfo{pages}{621–630}. \DOIprefix\doi{10.1145/1645953.1646033}.
\bibitem[{Clarke et~al.(2009)Clarke, Kolla, and Vechtomova}]{clarkeICTIR2009}
\bibinfo{author}{C.~L. Clarke}, \bibinfo{author}{M.~Kolla}, \bibinfo{author}{O.~Vechtomova},
\newblock \bibinfo{title}{An effectiveness measure for ambiguous and underspecified queries},
\newblock in: \bibinfo{booktitle}{Proceedings of the 2nd International Conference on Theory of Information Retrieval: Advances in Information Retrieval Theory}, ICTIR '09, \bibinfo{publisher}{Springer-Verlag}, \bibinfo{address}{Berlin, Heidelberg}, \bibinfo{year}{2009}, p. \bibinfo{pages}{188–199}. \DOIprefix\doi{10.1007/978-3-642-04417-5_17}.
\bibitem[{Clarke et~al.(2011)Clarke, Craswell, Soboroff, and Ashkan}]{clarkeWSDM2011}
\bibinfo{author}{C.~L. Clarke}, \bibinfo{author}{N.~Craswell}, \bibinfo{author}{I.~Soboroff}, \bibinfo{author}{A.~Ashkan},
\newblock \bibinfo{title}{A comparative analysis of cascade measures for novelty and diversity},
\newblock in: \bibinfo{booktitle}{Proceedings of the Fourth ACM International Conference on Web Search and Data Mining}, WSDM '11, \bibinfo{publisher}{Association for Computing Machinery}, \bibinfo{address}{New York, NY, USA}, \bibinfo{year}{2011}, p. \bibinfo{pages}{75–84}. \DOIprefix\doi{10.1145/1935826.1935847}.
\bibitem[{Ziegler et~al.(2005)Ziegler, McNee, Konstan, and Lausen}]{zieglerWWW2005}
\bibinfo{author}{C.-N. Ziegler}, \bibinfo{author}{S.~M. McNee}, \bibinfo{author}{J.~A. Konstan}, \bibinfo{author}{G.~Lausen},
\newblock \bibinfo{title}{Improving recommendation lists through topic diversification},
\newblock in: \bibinfo{booktitle}{Proceedings of the 14th International Conference on World Wide Web}, WWW '05, \bibinfo{publisher}{Association for Computing Machinery}, \bibinfo{address}{New York, NY, USA}, \bibinfo{year}{2005}, p. \bibinfo{pages}{22–32}. \DOIprefix\doi{10.1145/1060745.1060754}.
\bibitem[{Vargas and Castells(2011)}]{vargasRECSYS2011}
\bibinfo{author}{S.~Vargas}, \bibinfo{author}{P.~Castells},
\newblock \bibinfo{title}{Rank and relevance in novelty and diversity metrics for recommender systems},
\newblock in: \bibinfo{booktitle}{Proceedings of the Fifth ACM Conference on Recommender Systems}, RecSys '11, \bibinfo{publisher}{Association for Computing Machinery}, \bibinfo{address}{New York, NY, USA}, \bibinfo{year}{2011}, p. \bibinfo{pages}{109–116}. \DOIprefix\doi{10.1145/2043932.2043955}.
\bibitem[{Vargas(2014)}]{varagsSIGIR2014}
\bibinfo{author}{S.~Vargas},
\newblock \bibinfo{title}{Novelty and diversity enhancement and evaluation in recommender systems and information retrieval},
\newblock in: \bibinfo{booktitle}{Proceedings of the 37th International ACM SIGIR Conference on Research \& Development in Information Retrieval}, SIGIR '14, \bibinfo{publisher}{Association for Computing Machinery}, \bibinfo{address}{New York, NY, USA}, \bibinfo{year}{2014}, p. \bibinfo{pages}{1281}. \DOIprefix\doi{10.1145/2600428.2610382}.
\bibitem[{Zheng et~al.(2021)Zheng, Gao, Chen, Jin, and Li}]{zhengWWW2021}
\bibinfo{author}{Y.~Zheng}, \bibinfo{author}{C.~Gao}, \bibinfo{author}{L.~Chen}, \bibinfo{author}{D.~Jin}, \bibinfo{author}{Y.~Li},
\newblock \bibinfo{title}{Dgcn: Diversified recommendation with graph convolutional networks},
\newblock in: \bibinfo{booktitle}{Proceedings of the Web Conference 2021}, WWW '21, \bibinfo{publisher}{Association for Computing Machinery}, \bibinfo{address}{New York, NY, USA}, \bibinfo{year}{2021}, p. \bibinfo{pages}{401–412}. \DOIprefix\doi{10.1145/3442381.3449835}.
\bibitem[{Radlinski et~al.(2008)Radlinski, Kleinberg, and Joachims}]{radlinskiICML2008}
\bibinfo{author}{F.~Radlinski}, \bibinfo{author}{R.~Kleinberg}, \bibinfo{author}{T.~Joachims},
\newblock \bibinfo{title}{Learning diverse rankings with multi-armed bandits},
\newblock in: \bibinfo{booktitle}{Proceedings of the 25th International Conference on Machine Learning}, ICML '08, \bibinfo{publisher}{Association for Computing Machinery}, \bibinfo{address}{New York, NY, USA}, \bibinfo{year}{2008}, p. \bibinfo{pages}{784–791}. \DOIprefix\doi{10.1145/1390156.1390255}.
\bibitem[{Zhu et~al.(2014)Zhu, Lan, Guo, Cheng, and Niu}]{zhuSIGIR2014}
\bibinfo{author}{Y.~Zhu}, \bibinfo{author}{Y.~Lan}, \bibinfo{author}{J.~Guo}, \bibinfo{author}{X.~Cheng}, \bibinfo{author}{S.~Niu},
\newblock \bibinfo{title}{Learning for search result diversification},
\newblock in: \bibinfo{booktitle}{Proceedings of the 37th International ACM SIGIR Conference on Research \& Development in Information Retrieval}, SIGIR '14, \bibinfo{publisher}{Association for Computing Machinery}, \bibinfo{address}{New York, NY, USA}, \bibinfo{year}{2014}, p. \bibinfo{pages}{293–302}. \DOIprefix\doi{10.1145/2600428.2609634}.
\bibitem[{Desautels et~al.(2014)Desautels, Krause, and Burdick}]{desautelsJMLR2014}
\bibinfo{author}{T.~Desautels}, \bibinfo{author}{A.~Krause}, \bibinfo{author}{J.~W. Burdick},
\newblock \bibinfo{title}{Parallelizing exploration-exploitation tradeoffs in gaussian process bandit optimization},
\newblock \bibinfo{journal}{Journal of Machine Learning Research} \bibinfo{volume}{15} (\bibinfo{year}{2014}) \bibinfo{pages}{4053--4103}. \URLprefix \url{http://jmlr.org/papers/v15/desautels14a.html}.
\bibitem[{Contal et~al.(2013)Contal, Buffoni, Robicquet, and Vayatis}]{contalECMLPKDD2013}
\bibinfo{author}{E.~Contal}, \bibinfo{author}{D.~Buffoni}, \bibinfo{author}{A.~Robicquet}, \bibinfo{author}{N.~Vayatis},
\newblock \bibinfo{title}{Parallel gaussian process optimization with upper confidence bound and pure exploration},
\newblock in: \bibinfo{editor}{H.~Blockeel}, \bibinfo{editor}{K.~Kersting}, \bibinfo{editor}{S.~Nijssen}, \bibinfo{editor}{F.~{\v{Z}}elezn{\'y}} (Eds.), \bibinfo{booktitle}{Machine Learning and Knowledge Discovery in Databases}, \bibinfo{publisher}{Springer Berlin Heidelberg}, \bibinfo{address}{Berlin, Heidelberg}, \bibinfo{year}{2013}, pp. \bibinfo{pages}{225--240}.
\bibitem[{Ntoulas et~al.(2006)Ntoulas, Najork, Manasse, and Fetterly}]{ntoulasWWW2006}
\bibinfo{author}{A.~Ntoulas}, \bibinfo{author}{M.~Najork}, \bibinfo{author}{M.~Manasse}, \bibinfo{author}{D.~Fetterly},
\newblock \bibinfo{title}{Detecting spam web pages through content analysis},
\newblock in: \bibinfo{booktitle}{Proceedings of the 15th International Conference on World Wide Web}, WWW '06, \bibinfo{publisher}{Association for Computing Machinery}, \bibinfo{address}{New York, NY, USA}, \bibinfo{year}{2006}, p. \bibinfo{pages}{83–92}. \DOIprefix\doi{10.1145/1135777.1135794}.
\bibitem[{Cormack et~al.(2011)Cormack, Smucker, and Clarke}]{cormackIR2011}
\bibinfo{author}{G.~V. Cormack}, \bibinfo{author}{M.~D. Smucker}, \bibinfo{author}{C.~L.~A. Clarke},
\newblock \bibinfo{title}{Efficient and effective spam filtering and re-ranking for large web datasets},
\newblock \bibinfo{journal}{Inf. Retr.} \bibinfo{volume}{14} (\bibinfo{year}{2011}) \bibinfo{pages}{441–465}. \DOIprefix\doi{10.1007/s10791-011-9162-z}.
\bibitem[{Bendersky et~al.(2011)Bendersky, Croft, and Diao}]{benderskyWSDM2011}
\bibinfo{author}{M.~Bendersky}, \bibinfo{author}{W.~B. Croft}, \bibinfo{author}{Y.~Diao},
\newblock \bibinfo{title}{Quality-biased ranking of web documents},
\newblock in: \bibinfo{booktitle}{Proceedings of the Fourth ACM International Conference on Web Search and Data Mining}, WSDM '11, \bibinfo{publisher}{Association for Computing Machinery}, \bibinfo{address}{New York, NY, USA}, \bibinfo{year}{2011}, p. \bibinfo{pages}{95–104}. \DOIprefix\doi{10.1145/1935826.1935849}.
\bibitem[{Brin and Page(1998)}]{brinCISDN1998}
\bibinfo{author}{S.~Brin}, \bibinfo{author}{L.~Page},
\newblock \bibinfo{title}{The anatomy of a large-scale hypertextual web search engine},
\newblock \bibinfo{journal}{Computer Networks and ISDN Systems} \bibinfo{volume}{30} (\bibinfo{year}{1998}) \bibinfo{pages}{107--117}. \URLprefix \url{https://www.sciencedirect.com/science/article/pii/S016975529800110X}, \bibinfo{note}{proceedings of the Seventh International World Wide Web Conference}.
\bibitem[{Lin et~al.(2021)Lin, Hilton, and Evans}]{Lin2021TruthfulQAMH}
\bibinfo{author}{S.~C. Lin}, \bibinfo{author}{J.~Hilton}, \bibinfo{author}{O.~Evans},
\newblock \bibinfo{title}{Truthfulqa: Measuring how models mimic human falsehoods},
\newblock in: \bibinfo{booktitle}{Annual Meeting of the Association for Computational Linguistics}, \bibinfo{year}{2021}. \URLprefix \url{https://api.semanticscholar.org/CorpusID:237532606}.
\bibitem[{Rafailov et~al.(2023)Rafailov, Sharma, Mitchell, Manning, Ermon, and Finn}]{rafailovNEURIPS2023}
\bibinfo{author}{R.~Rafailov}, \bibinfo{author}{A.~Sharma}, \bibinfo{author}{E.~Mitchell}, \bibinfo{author}{C.~D. Manning}, \bibinfo{author}{S.~Ermon}, \bibinfo{author}{C.~Finn},
\newblock \bibinfo{title}{Direct preference optimization: Your language model is secretly a reward model},
\newblock in: \bibinfo{editor}{A.~Oh}, \bibinfo{editor}{T.~Neumann}, \bibinfo{editor}{A.~Globerson}, \bibinfo{editor}{K.~Saenko}, \bibinfo{editor}{M.~Hardt}, \bibinfo{editor}{S.~Levine} (Eds.), \bibinfo{booktitle}{Advances in Neural Information Processing Systems}, volume~\bibinfo{volume}{36}, \bibinfo{publisher}{Curran Associates, Inc.}, \bibinfo{year}{2023}, pp. \bibinfo{pages}{53728--53741}. \URLprefix \url{https://proceedings.neurips.cc/paper_files/paper/2023/file/a85b405ed65c6477a4fe8302b5e06ce7-Paper-Conference.pdf}.
\bibitem[{{Jiang} et~al.(2023){Jiang}, {Sablayrolles}, {Mensch}, {Bamford}, {Singh Chaplot}, {de las Casas}, {Bressand}, {Lengyel}, {Lample}, {Saulnier}, {Renard Lavaud}, {Lachaux}, {Stock}, {Le Scao}, {Lavril}, {Wang}, {Lacroix}, and {El Sayed}}]{jiang2023arXiv}
\bibinfo{author}{A.~Q. {Jiang}}, \bibinfo{author}{A.~{Sablayrolles}}, \bibinfo{author}{A.~{Mensch}}, \bibinfo{author}{C.~{Bamford}}, \bibinfo{author}{D.~{Singh Chaplot}}, \bibinfo{author}{D.~{de las Casas}}, \bibinfo{author}{F.~{Bressand}}, \bibinfo{author}{G.~{Lengyel}}, \bibinfo{author}{G.~{Lample}}, \bibinfo{author}{L.~{Saulnier}}, \bibinfo{author}{L.~{Renard Lavaud}}, \bibinfo{author}{M.-A. {Lachaux}}, \bibinfo{author}{P.~{Stock}}, \bibinfo{author}{T.~{Le Scao}}, \bibinfo{author}{T.~{Lavril}}, \bibinfo{author}{T.~{Wang}}, \bibinfo{author}{T.~{Lacroix}}, \bibinfo{author}{W.~{El Sayed}},
\newblock \bibinfo{title}{{Mistral 7B}},
\newblock \bibinfo{journal}{arXiv e-prints}  (\bibinfo{year}{2023}) \bibinfo{pages}{arXiv:2310.06825}. \DOIprefix\doi{10.48550/arXiv.2310.06825}. \href{http://arxiv.org/abs/2310.06825}{{\tt arXiv:2310.06825}}.
\bibitem[{Liu et~al.(2022)Liu, Tam, Muqeeth, Mohta, Huang, Bansal, and Raffel}]{liuNEURIPS2022}
\bibinfo{author}{H.~Liu}, \bibinfo{author}{D.~Tam}, \bibinfo{author}{M.~Muqeeth}, \bibinfo{author}{J.~Mohta}, \bibinfo{author}{T.~Huang}, \bibinfo{author}{M.~Bansal}, \bibinfo{author}{C.~A. Raffel},
\newblock \bibinfo{title}{Few-shot parameter-efficient fine-tuning is better and cheaper than in-context learning},
\newblock in: \bibinfo{editor}{S.~Koyejo}, \bibinfo{editor}{S.~Mohamed}, \bibinfo{editor}{A.~Agarwal}, \bibinfo{editor}{D.~Belgrave}, \bibinfo{editor}{K.~Cho}, \bibinfo{editor}{A.~Oh} (Eds.), \bibinfo{booktitle}{Advances in Neural Information Processing Systems}, volume~\bibinfo{volume}{35}, \bibinfo{publisher}{Curran Associates, Inc.}, \bibinfo{year}{2022}, pp. \bibinfo{pages}{1950--1965}. \URLprefix \url{https://proceedings.neurips.cc/paper_files/paper/2022/file/0cde695b83bd186c1fd456302888454c-Paper-Conference.pdf}.
\bibitem[{Li and Chan(2013)}]{liVLDB2013}
\bibinfo{author}{L.~Li}, \bibinfo{author}{C.-Y. Chan},
\newblock \bibinfo{title}{Efficient indexing for diverse query results},
\newblock \bibinfo{journal}{Proc. VLDB Endow.} \bibinfo{volume}{6} (\bibinfo{year}{2013}) \bibinfo{pages}{745–756}. \DOIprefix\doi{10.14778/2536360.2536373}.

\end{thebibliography}

\end{document}